\documentclass[aps,prd,preprint,tightenlines,nofootinbib,showpacs,fixfloat]{revtex4}
\usepackage{epsfig}
\usepackage{graphicx}
\usepackage{dcolumn}
\usepackage{bm}
\usepackage{overpic}
\usepackage{subfigure}
\usepackage{float}
\usepackage{color}

\def \bea{\begin{eqnarray}}
\def \beq{\begin{equation}}

\def \ec{\eta_c}
\def \eea{\end{eqnarray}}
\def \eeq{\end{equation}}
\def \ege1{E(\gamma_{\rm E1})}

\def \gme1{\gamma_{\rm E1}}
\def \hc{h_c}
\def \jp{J/\psi}

\def \ks{K^0_S}

\def \pp{\psi(3686)}

\def \pz{\pi^0}

\begin{document}


\title{\boldmath Study of $\psi(3686)\to\pi^0 h_c, h_c\to\gamma\eta_c$ via $\eta_c$ exclusive decays}

\author{
     {\small
     M.~Ablikim$^{1}$, M.~N.~Achasov$^{5}$, O.~Albayrak$^{3}$, D.~J.~Ambrose$^{39}$, F.~F.~An$^{1}$, Q.~An$^{40}$, J.~Z.~Bai$^{1}$, Y.~Ban$^{27}$, J.~Becker$^{2}$, J.~V.~Bennett$^{17}$
% ...
\\
  \vspace{0.2cm}
 (BESIII Collaboration)\\
\vspace{0.2cm} {\it
$^{1}$ Institute of High Energy Physics, Beijing 100049, P. R. China\\
$^{2}$ Bochum Ruhr-University, 44780 Bochum, Germany\\
$^{3}$ Carnegie Mellon University, Pittsburgh, PA 15213, USA\\
% ...
$^{5}$ G.I. Budker Institute of Nuclear Physics SB RAS (BINP), Novosibirsk 630090, Russia\\
% ...
$^{17}$ Indiana University, Bloomington, Indiana 47405, USA\\
% ...
$^{27}$ Peking University, Beijing 100871, P. R. China\\
% ...
$^{39}$ University of Rochester, Rochester, New York 14627, USA\\
$^{40}$ University of Science and Technology of China, Hefei 230026, P. R. China\\
% ...
}
}}
\date{\today}

\begin{abstract}
The process $\psi(3686) \to \pi^0 h_c, h_c \to \gamma \eta_c$ has been studied
with a data sample of $106 \pm 4$ million $\psi(3686)$ events collected with the BESIII detector
at the BEPCII storage ring.  The mass and width of the $P$-wave charmonium spin-singlet
state $h_c(^1P_1)$ are determined by simultaneously fitting distributions of the $\pi^0$
recoil mass for 16 exclusive $\eta_c$ decay modes.  The results,
$M(\hc) = 3525.31 \pm 0.11~{\rm (stat.)} \pm 0.14~{\rm (syst.)}$\,MeV/$c^2$ and
$\Gamma(\hc) = 0.70 \pm 0.28 \pm 0.22$\,MeV, are consistent with and more precise
than previous measurements.  We also determine the branching ratios for the 16 exclusive
$\eta_c$  decay modes, five of which have not been measured previously.  New measurements
of the $\eta_c$ line-shape parameters in the $E1$ transition $h_c\to\gamma\eta_c$ are
made by selecting candidates in the $h_c$ signal sample and simultaneously fitting the
hadronic mass spectra for the 16 $\eta_c$ decay channels.  The resulting $\eta_c$ mass
and width values are $M(\eta_c) = 2984.49 \pm 1.16 \pm 0.52$\,MeV/$c^2$
and $\Gamma(\eta_c) = 36.4 \pm 3.2 \pm 1.7$\,MeV.
\end{abstract}

\maketitle
\section{\bf INTRODUCTION}
\bigskip

Studies of charmonium states have played an important role in
understanding Quantum Chromodynamics (QCD) because of their relative
immunity from complications like relativistic effects and the large
value of the strong coupling constant $\alpha_s$.  In the QCD
potential model~\cite{Eichten:1978tg}, the spin-independent
one-gluon exchange part of the $c\bar{c}$ interaction has been
defined quite well by existing experimental data.  The spin
dependence of the $c\bar{c}$ potential is not as well understood.
Until recently, the only well-measured hyperfine splitting was that
for the $1S$ states of charmonium, $\Delta M_{hf}(1 S )= M (J/\psi )
- M (\eta_c) = 116 \pm 1$\,MeV/$c^2$~\cite{ref:PDG_2012}.  In the past
several years Belle~\cite{Choi:2002na}, CLEO~\cite{etac_prime_cleo}, BaBar~\cite{etac_prime_babar}, and BESIII~\cite{Ablikim:2012sf} have succeeded in identifying
$\eta_c(2S)$ and have measured $\Delta M_{hf}(2S)=M (\psi(3686)) -
M(\eta_c(2S))=47 \pm 1$\,MeV/$c^2$.

Of the charmonium states below $D\bar{D}$ threshold, the
$h_c(1^{1}P_{1})$ is experimentally the least accessible because it
cannot be produced directly in $e^+e^-$ annihilation or in the
electric-dipole transition of a $J^{PC} = 1^{--}$ charmonium state.
Limited statistics and photon-detection challenges also were major
obstacles to the observation of $h_c$ in charmonium transitions. The
precise measurement of $h_c$ properties is important because a
comparison of its mass with the masses of the $3P$ states
($\chi_{cJ}$) provides much-needed information about the spin
dependence of the $c\bar{c}$ interaction.
According to QCD potential models, the $c\bar{c}$ interaction in a
charmonium meson can be described with a potential that includes a
Lorentz scalar confinement term and a vector Coulombic term arising
from one-gluon exchange between the quark and the antiquark.  The
scalar confining potential makes no contribution to the hyperfine
interaction and the Coulombic vector potential produces hyperfine
splitting only for $S$ states.  This leads to the prediction of the
hyperfine or triplet-singlet splitting in the $P$ states of $M_{hf}
\equiv\langle M(1^3P)\rangle-M(1^1P_1)\simeq 0$, where $\langle M(1^3P)\rangle$
is the spin-weighted centroid mass of the triplet $^3P_J$ states~\cite
{swanson,ref:E835hc,ref:cleohc08}.

The first evidence of the $h_c$ state was reported by the Fermilab
E760 experiment~\cite{ref:E760hc} and was based on the process
$p\bar{p}\to \pi^0J/\psi$.  This result was subsequently excluded by
the successor experiment E835~\cite{ref:E835hc}, which investigated
the same reaction with a larger data sample.   E835 also studied
$p\bar{p}\to h_c\to\gamma\eta_c$, in this case finding an $h_c$
signal.  Soon after this the CLEO collaboration observed the $h_c$
and measured its mass~\cite{ref:cleohc05,ref:cleohc08} by studying
the decay chain $\psi(3686)\to\pi^0 h_c, h_c\to\gamma\eta_c$ in
$e^+e^-$ collisions.  CLEO subsequently presented evidence for $h_c$
decays to multi-pion final states~\cite{ref:cleohc09}. Recently, the
BESIII collaboration used inclusive methods to make the first
measurements of the absolute branching ratios
$\mathcal{B}(\psi(3686)\to\pi^{0}h_c)=(8.4\pm1.3\pm1.0)\times10^{-4}$
and $\displaystyle
\mathcal{B}(h_c\to\gamma\eta_c)=(54.3\pm6.7\pm5.2)\%$~\cite{ref:bes3hc10}.
CLEO has confirmed the BESIII results~\cite{Ge:2011kq} and also
observed $h_c$ in $e^+e^-\to \pi^+\pi^- h_c$ at $\sqrt{s}
=4170$\,MeV, demonstrating a new prolific source of
$h_c$~\cite{CLEO:2011aa}.

$\eta_c(1S)$ is the lowest-lying $S$-wave spin-singlet charmonium
state. Although it has been known for about thirty
years~\cite{Himel:1980dj}, its resonant parameters are still
interesting.  For a long time, measurements of the $\eta_c$ width
from B-factories and from charmonium transitions were inconsistent
\cite{ref:PDG_2012}.  The discrepancies can be attributed to poor
statistics and inadequate consideration of interference between
$\eta_c$ decays and non-resonant backgrounds. Besides, the $\eta_c$ line shape also
could be distorted by photon energy dependence in the $M1$ (or $E1$) transition,
which will affect the resonant-parameter measurements. Recent studies by
Belle, Babar, CLEO, and
BESIII~\cite{Vinokurova:2011dy,delAmoSanchez:2011bt,Mitchell:2008aa,BESIII:2011ab},
with large data samples and careful consideration of interference,
obtained similar $\eta_c$ width and mass results in
two-photon-fusion production and $\psi(3686)$ decays. The $h_c \to
\gamma \eta_c$ transition can provide a new laboratory to study
$\eta_c$ properties. The $\eta_c$ line shape in the $E1$ transition
$h_c\to\gamma\eta_c$ should not be as distorted as in other charmonium decays, because
non-resonant interfering backgrounds to the dominant transition are
small.

In this paper, we report new measurements of the mass and width of
the $h_c$ and $\eta_c$, and of the branching ratios
$\mathcal{B}_1(\psi(3686)\to\pi^{0}h_c)\times
\mathcal{B}_2(h_c\to\gamma\eta_c)\times \mathcal{B}_3(\eta_c\to
X_i)$ and $\mathcal{B}_3(\eta_c\to X_i)$, via the sequential process
$\psi(3686) \to \pi^0 h_c,~h_c \to \gamma\eta_c,~\eta_c \to X_i$. In
this reaction $X_i$ signifies 16 exclusive hadronic final states: $p
\bar{p}$, $2(\pi^+ \pi^-)$, $2(K^+ K^-)$, $K^+ K^- \pi^+ \pi^-$, $p
\bar{p} \pi^+ \pi^-$, $3(\pi^+ \pi^-)$, $K^+ K^- 2(\pi^+ \pi^-)$,
$K^+ K^- \pi^0$, $p \bar{p}\pi^0$, $\ks K^\pm \pi^\mp$, $\ks K^\pm
\pi^\mp \pi^\pm \pi^\mp$, $\pi^+ \pi^- \eta$, $K^+ K^- \eta$,
$2(\pi^+ \pi^-) \eta$, $\pi^+ \pi^- \pi^0 \pi^0$, and $2(\pi^+
\pi^-) \pi^0 \pi^0$. Here $K_S^0$ is reconstructed in its
$\pi^+\pi^-$ decays, and $\eta$ in its $\gamma \gamma$ final state.
The data sample of $\psi(3686)$ events was collected with the BESIII
detector at the BEPCII $e^+e^-$ storage ring.

The remainder of this paper is structured as follows: Sect.~\ref{sec:expt}
describes the experiment and data sample; Sect.~\ref{sec:sel} presents
the event selection and background analysis; Sect.~\ref{sec:meas} discusses
the extraction of $h_c$ and $\eta_c$ results;  Sect.~\ref{sec:sys} describes
the estimation of systematic uncertainties; and Sec.~\ref{sec:sum} provides a
summary and discussion of the results.

\section{\bf Experiment and Data Sample\label{sec:expt}}
\bigskip

BEPCII is a two-ring $e^+e^-$ collider designed for a peak
luminosity of $10^{33}$\,cm$^{-2}s^{-1}$ at a beam current of
0.93\,A per beam. The cylindrical core of the BESIII detector
consists of a helium-gas-based main drift chamber for
charged-particle tracking and particle identification by d$E$/d$x$,
a plastic scintillator time-of-flight system for additional
particle identification, and a 6240-crystal CsI(Tl) Electromagnetic
Calorimeter~(EMC) for electron identification and photon detection.
These components are all enclosed in a superconducting solenoidal
magnet providing a 1.0-T magnetic field.  The solenoid is supported
by an octagonal flux-return yoke with resistive-plate-counter muon
detector modules interleaved with steel.  The geometrical
acceptance for charged tracks and photons is $93\%$ of $4\pi$, and
the resolutions for charged-track momentum and photon energy at
1\,GeV are $0.5\%$ and $2.5\%$, respectively.  More details on the
features and capabilities of BESIII are provided in
Ref.~\cite{ref:bes3}.

The data sample for this analysis consists of 156.4\,pb$^{-1}$ of $e^+e^-$ annihilation
data collected at a center-of-mass energy of 3.686\,GeV, the peak of the $\psi(3686)$
resonance.  By measuring the production of multihadronic events we determine
the number of $\psi(3686)$ decays in the sample to be $(1.06 \pm 0.04) \times 10^8$,
where the uncertainty is dominated by systematics~\cite{ref:psiptotnumber}.  An
additional 42\,$\rm pb^{-1}$ of data were collected at a center-of-mass energy
of 3.65\,GeV to determine non-resonant continuum background contributions.

The optimization of the event selection and the estimation of physics backgrounds
are performed with simulated Monte Carlo~(MC) samples.  A
{\tt GEANT4}-based~\cite{Agostinelli:2002hh,Allison:2006ve} detector simulation
package is used to model the detector response.  Signal and background processes
are generated with specialized models that have been packaged and customized
for BESIII~\cite{ref:bes3gen}.  The $\psi(3686)$ resonance is generated by
{\tt KKMC}~\cite{ref:kkmc}, and {\tt EvtGen}~\cite{ref:evtgen} is used to model
events for $\psi(3686)\to\pi^{0}h_{c}$ and for exclusive backgrounds in
$\psi(3686)$ decays. An inclusive sample (100 million events) is used to simulate
hadronic background processes.  Known $\psi(3686)$ decay modes are generated with
{\tt EvtGen}, using branching ratios set to world-average values~\cite{ref:PDG_2012}.
The remaining $\psi(3686)$ decay modes are generated by
{\tt LUNDCHARM}~\cite{ref:bes3gen}, which is based on
{\tt JETSET}~\cite{Sjostrand:1993yb}
and tuned for the charm-energy region.  The decays $\psi(3686)\to \pi^0 h_c$ are
excluded from this sample.

The $\psi(3686)\to\pi^{0}h_{c}$ events are generated with an $h_c$ mass of
$3525.28$\,MeV/$c^2$  and a width equal to that of the $\chi_{c1}$ (0.9\,MeV).  The
$E1$ transition $h_c\to\gamma\eta_c$ is generated with an angular distribution in
the $h_c$ rest frame of $1 +\cos^2\theta^*$, where $\theta^*$ is the angle
of the E1 photon with respect to the beam direction in the $h_c$ rest frame.
Multi-body $\eta_c$ decays are generated according to phase space.

\section{\bf Event selection and background analysis\label{sec:sel}}
\bigskip

For $\psi(3686)\to\pi^0 h_c$, $h_c\to\gamma\eta_c$, the expected $\pi^0$ momentum
is $P_{\pi^0} \simeq 84$\,MeV/$c$, and the $E1$ transition photon
emitted in $\hc \to \gamma \ec$ has an expected energy of $\ege1
\simeq 503$\,MeV in the $\hc$ rest frame.
Therefore, the signal candidates should have one $E1$ photon candidate with energy in the
expected region $450\,\rm{MeV}<\ege1<550\,\rm{MeV}$ and one $\pi^0$ candidate with recoil mass in the
region $(3480,3570)\,\rm{MeV/c^2}$.
For the selected candidates, we fit the distribution
of $\pi^0$ recoil mass for the full event sample to give the results for the $h_c$
resonant parameters and signal yields.

Charged tracks in BESIII are reconstructed from main drift chamber hits within a
polar-angle ($\theta$) acceptance range of $|\cos\theta| < 0.93$. To
optimize the momentum measurement, we require that these tracks be
reconstructed to pass within 10\,cm of the interaction point in the
beam direction and within 1\,cm in the plane perpendicular to the
beam.  Tracks used in reconstructing $K^0_S$ decays are exempted
from these requirements.

A vertex fit constrains charged tracks to a common production vertex, which
is updated on a run-by-run basis.  For each charged track, time-of-flight and
d$E$/d$x$ information is combined to compute particle identification (PID)
confidence levels for the pion, kaon, and proton hypotheses.  The track is
assigned to the particle type with the highest confidence level.

Electromagnetic showers are reconstructed by clustering EMC crystal
energies.  Efficiency and energy resolution are improved by including
energy deposits in nearby time-of-flight counters.  A photon candidate is defined
as a shower with an energy deposit of at least 25\,MeV in the ``barrel''
region~($|\cos \theta| < 0.8$), or of at least 50\,MeV in the
``end-cap'' region~($0.86 < |\cos\theta| < 0.92$).  Showers at angles
intermediate between the barrel and the end-cap are not well measured and
are rejected.  An additional requirement on the EMC hit timing suppresses electronic
noise and energy deposits unrelated to the event.

A candidate $\pi^0$($\eta$) is reconstructed from pairs of photons
with an invariant mass in the range
$|M_{\gamma\gamma}-m_{\pi^0}|<15\,\rm MeV/c^2$
($|M_{\gamma\gamma}-m_{\eta}|<15\,\rm MeV/c^2$)~\cite{ref:PDG_2012}.
A one-constraint~(1-C) kinematic fit is performed to improve the
energy resolution, with the $M(\gamma\gamma)$ constrained to the
known $\pi^0$($\eta$) mass.

We reconstruct $K^0_S\to\pi^+\pi^-$ candidates using pairs of
oppositely charged tracks with an invariant mass in the range
$|M_{\pi\pi}-m_{K^0_S}| < 20\,\rm MeV/c^2$, where $m_{K^0_S}$ is the
known $K^0_S$ mass~\cite{ref:PDG_2012}. To reject random $\pi^+\pi^-$ combinations, a
secondary-vertex fitting algorithm is employed to impose the
kinematic constraint between the production and decay
vertices~\cite{ref:ks0-reconstruction}.  Accepted $K^0_S$ candidates
are required to have a decay length of at least twice the vertex
resolution.

The $\eta_c$ candidate is reconstructed in 16 exclusive decay modes,
and the event is accepted or rejected based on consistency with the
$h_c\to\gamma\eta_c$ hypothesis.  Specifically, the reconstructed
mass $M(\eta_c)$ is required to be between 2.900\,GeV/$c^2$ and
3.050\,GeV/$c^2$, and the transition-photon energy is required to be
between 0.450\,GeV and 0.550\,GeV.
Events passing this selection are subjected to a 4 constraint (4-C) kinematic fit to take advantage of energy-momentum conservation between the initial state ($e^+e^-$ beams) and the final state ($\eta_c+E1~\rm{photon}+\pi^0$).
Because of differing signal/background characteristics, we individually optimize requirements on
$\chi^2_{4C}$, the $\chi^2$ of the 4-C fit, for the 16 $\eta_c$
channels.  If multiple $\eta_c$ candidates are found in an event,
the one with the smallest value of $\chi^2 = \chi^2_{\rm
4C}+\chi^2_{\rm 1C}+\chi^2_{\rm pid}+\chi^2_{\rm vertex}$ is
accepted, where $\chi^2_{\rm 1C}$ is the $\chi^2$ of the 1-C fit of
the $\pi^0$($\eta$), $\chi^2_{\rm pid}$ is the PID $\chi^2$
summation for all charged tracks included in the $h_c$ candidate,
and $\chi^2_{\rm vertex}$ is the $\chi^2$ of the $K_S^0$ vertex fit.
If there is no $\pi^0/\eta$ ($K_S^0$) in an event, the corresponding
$\chi^2_{\rm 1C}$ ($\chi^2_{\rm vertex}$) is set to zero.

Based on studies of the inclusive MC sample, we identified several
background processes with potential to reduce the precision of measurements
made with specific $\eta_c$ exclusive channels because of sizable low-energy
$\pi^0$ production.  The processes and suppression procedures are as follows:

\begin{itemize}

\item {\it $\pp \to \pi^+ \pi^- \jp$}

The mass $M_X$ of the system recoiling against the $\pi^+\pi^-$
in $\pp \to \pi^+ \pi^- X$ is calculated and the candidate is rejected if
$M_X$ is within $\pm 12$\,MeV/$c^2$ of the known $J/\psi$ mass.

\item {\it $\pp \to \pz \pz \jp$}

The mass $M_X$ of the system recoiling against the $\pi^0\pi^0$
in $\pp \to \pz \pz X$ is calculated and the candidate is rejected if
$M_X$ is within $\pm 15$\,MeV/$c^2$ of the known $J/\psi$ mass
for all $\eta_c$ final states except $\pi^+\pi^-\pi^+\pi^-\pi^0\pi^0$.
For this mode the lower $\pi^0$ momentum leads to recoil masses near
3.1\,GeV/$c^2$, so the exclusion window is narrowed to $\pm 10$\,MeV/$c^2$.

\item {\it $\pp \to \gamma \chi_{c2}$}

A candidate is rejected if it includes a $\pi^0$ for which either daughter photon has
an energy within $\pm5$\,MeV of that expected for the $\psi(3686)$ radiative
transition to $\chi_{c2}$ (128\,MeV).

\item {\it $E1$ photon candidates that are $\pz$ decay products}

A candidate is rejected if its $E1$ photon can be combined with another photon
in the event to form a $\pi^0$ within a mass window of $\pm 10$\,MeV/$c^2$.

\item {\it $\pi^0$ candidates that are from $\eta\to\pi^+\pi^-\pi^0$}

Masses $M(\pi^+\pi^-\pi^0)$ are calculated for all possible combinations in the
event and the candidate is rejected if any combination has a mass within
$\pm 15$\,MeV/$c^2$ of the known $\eta$ mass.

\end{itemize}

Decisions about whether to apply a requirement to a particular
$\eta_c$ mode and the optimization of the $\chi^2_{4C}$ and PID
requirements were made on a channel-by-channel basis.  The
figure-of-merit used was $\mathcal{S}=N_{S}/\sqrt{N_{S}+N_{B}}$,
where $N_{S}$ is the number of signal and $N_{B}$ the number of
background candidates.  Particle data group (PDG) values~\cite{ref:PDG_2012} are used for
the input $\eta_c$ branching ratios, and for channels not tabulated
by the PDG we estimate branching ratios based on conjugate channels
or other similar modes. The optimized selection criteria are listed
in Table~\ref{tab:excselection}, in which the $N(p),~N(\pi)~{\rm
and}~N(K)$ denote the numbers of identified protons, pions and kaons
in an event.

\begin{table*}[htbp]
\small
\begin{center}
\caption{Event-selection requirements for each exclusive channel.
\label{tab:excselection}}
\vspace{0.2cm}
\resizebox{!}{3.0cm}{
\begin{tabular}{l|c|c|c|c|c|c|c}
\hline \hline
Mode        & $\chi^2_{\rm~4C}$ &  PID & $\pi^+\pi^- J/\psi$  veto & $\pi^0\pi^0 J/\psi$ veto  & $\gamma\chi_{c2}$ veto  & $\pi^0$ veto  for $E1$ photon  & $\eta\to\pi^+\pi^-\pi^0$ veto   \\ \hline

$p \bar{p}$                                                & 30 &  $N(p)\geq1$               & no & no & yes & no & no   \\
$\pi^+ \pi^- \pi^+ \pi^-$                                  & 60 &  $N(\pi)\geq3$             & yes & yes & yes & yes & yes \\
$K^+ K^- K^+ K^-$                                          & 60 &  $N(K)\geq3$               & no & no & no & yes & no    \\
$K^+ K^- \pi^+ \pi^-$                                      & 40 &  $N(K)\geq2,N(\pi)\geq0$   & yes & yes & yes & yes & yes  \\
$p \bar{p} \pi^+ \pi^-$                                    & 30 &  $N(p)\geq2, N(\pi)\geq0$  & yes & yes & yes & yes & yes \\
$\pi^+ \pi^- \pi^+ \pi^- \pi^- \pi^-$                      & 50 &  $N(\pi)\geq4$             & yes & yes & no & yes & yes  \\
$K^+ K^- \pi^+ \pi^- \pi^- \pi^-$                          & 70 &  $N(K)\geq2, N(\pi)\geq2$  & yes & no & no & no & no   \\
$K^+ K^- \pi^0$                                            & 50 &  $N(K)\geq1$               & no & yes & no & no & no  \\
$p \bar{p} \pi^0$                                          & 40 &  $N(p)\geq1$               & no & yes & yes & yes & no  \\
$\ks K^\pm \pi^\mp$                                        & 70 &  $-$                       & no & no & no & no & yes   \\
$\ks K^\pm \pi^\mp \pi^\pm \pi^\mp $                       & 50 &  $-$                       & no & no & yes & no & no   \\
$\pi^+ \pi^- \eta$                                         & 50 &  $-$                       & no & no & no & yes & no   \\
$K^+ K^- \eta $                                            & 70 &  $N(K)\geq1$               & no & no & yes & yes & no   \\
$\pi^+ \pi^- \pi^+ \pi^- \eta$                             & 30 &  $-$                       & yes & no & no & yes & no  \\
$\pi^+ \pi^- \pi^0 \pi^0$                                  & 40 &  $-$                       & yes & yes &yes & yes & yes   \\
$\pi^+ \pi^- \pi^+ \pi^- \pi^0 \pi^0$                      & 60 &  $-$                       & yes & yes & no & yes & no   \\ \hline \hline

\end{tabular}
}
\end{center}
\end{table*}

\begin{figure}[tbhp]
\begin{center}
\epsfig{file=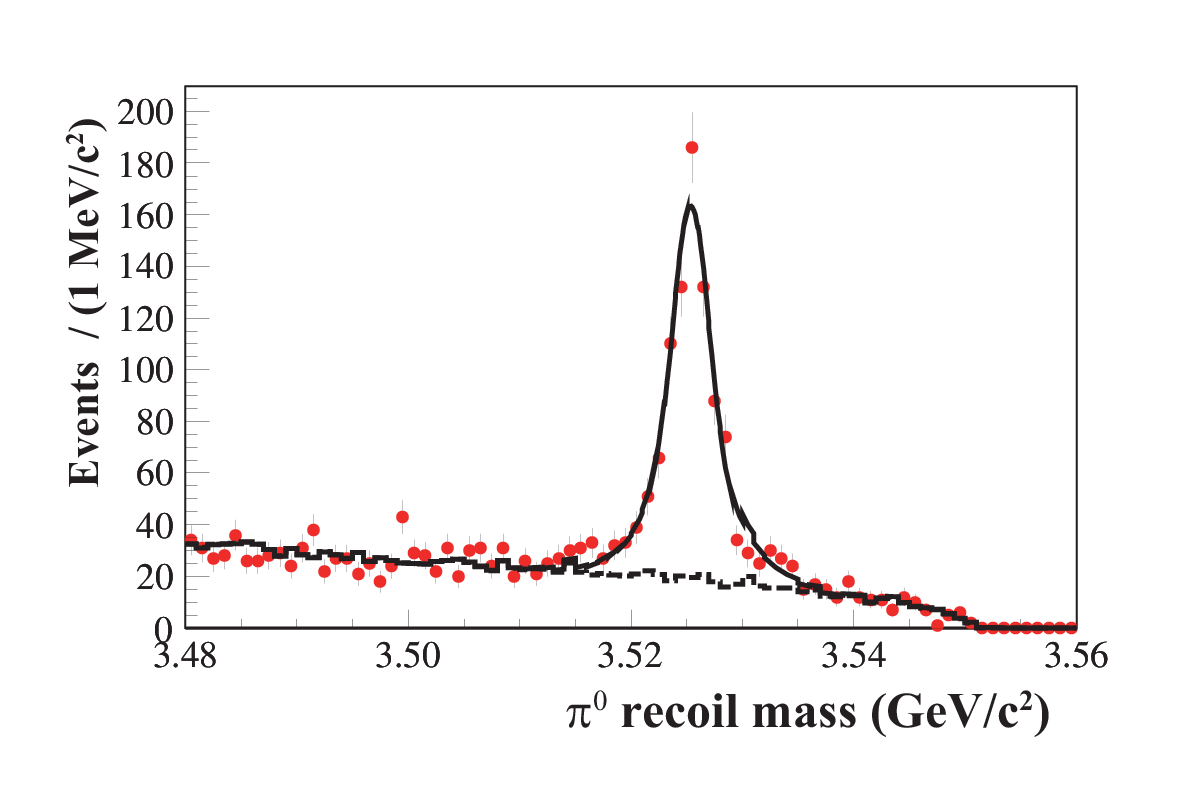,width=14cm} \caption{The $\pi^0$
recoil mass spectrum in $\psi(3686)\to\pi^0{h}_c,
{h}_c\to\gamma\eta_c$, $\eta_c \to X_i$ summed over the 16 final
states $X_i$.  The dots with error bars represent the $\pi^0$ recoil
mass spectrum in data. The solid line shows the total fit function
and the dashed line is the background component of the fit.
\label{fig:fithcpi0tot}}
\end{center}
\end{figure}

The $\pi^{0}$ recoil mass spectra for events passing these requirements show
clear $h_c$ signals in the expected range, as can be seen in Fig.~\ref{fig:fithcpi0tot}.
No peaking backgrounds in the signal region are found in the 100-million-event
inclusive MC sample, in the continuum data sample taken at $\sqrt{s}=3.65$\,GeV,
or in $\eta_c$-candidate-mass side-band distributions.

\section{\bf EXTRACTION OF YIELDS AND RESONANCE PARAMETERS\label{sec:meas}}

We obtain the $h_c$ mass, width and branching ratios from
simultaneous fits to the $\pi^{0}$ recoil mass distributions for the
16 exclusive $\eta_c$ decay modes. Here only 1-C kinematic fits with
$\pi^0$ mass hypothesis are used to improve the energy resolution.
The 4C-fits used in event selection are not used in the $\pi^0$
recoil mass reconstruction, because the energy resolution of the
signal $\pi^0$ in 4C-fits is not as good as in the 1C-fits,
according to a MC study. From the same data sample we also determine
the $\eta_c$ resonant parameters by fitting the 16 invariant-mass
spectra of the hadronic system accompanying the transition photon in
$\hc \to \gamma\ec$.

\subsection{\bf Fitting the {\boldmath $h_c$} signal \label{hc_fit}}
\bigskip

\begin{figure}[tbhp]
\centering
\includegraphics[width=16cm]{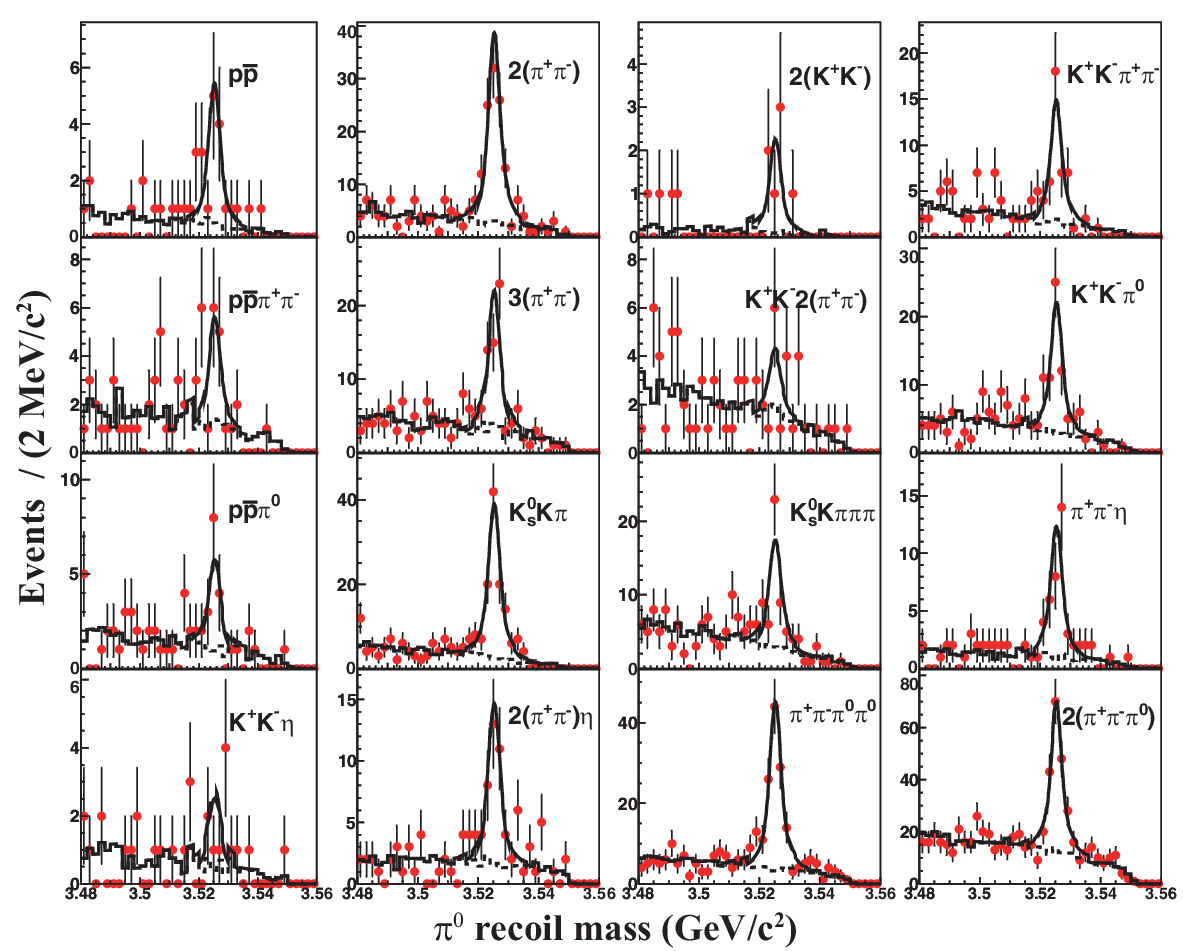}
\caption{The simultaneously fitted $\pi^0$ recoil mass spectra in
$\psi(3686)\to\pi^0{h}_c, {h}_c\to\gamma\eta_c$, $\eta_c \to X_i$ for the 16 final states $X_i$.}
\label{fig:fithcpi0sim}
\end{figure}

To extract the $h_c$ resonant parameters and the yield for each
$\eta_c$ decay channel, the 16 $\pi^{0}$ recoil mass distributions are
fitted simultaneously with a binned maximum likelihood method.  A
Breit-Wigner function convolved with the instrumental resolution is
used to describe the signal shape.  An efficiency correction is not needed
because of the small $h_c$ width and the good $\pi^0$ mass resolution.
The resolution function is channel-dependent
and is obtained from MC simulation.  The parameters $M(h_c)$ and
$\Gamma(h_c)$ of the Breit-Wigner function are constrained to be the same
for all 16 channels, which is essential for the decay modes with low
statistics.  For the recoil mass fit to each channel, the background shape is
obtained from the $\eta_c$ mass side bands (2300$-$2700, 3070$-$3200\,MeV/$c^2$), and the signal and the background
normalizations for each mode are allowed to float.
The summed and mode-by-mode fit results are shown in Figs.~\ref{fig:fithcpi0tot}
and \ref{fig:fithcpi0sim}, respectively.  The $\chi^2$ per
degree of freedom for this fit is 1.60, where sparsely populated bins are combined
so that there are at least seven counts per bin in the $\chi^2$ calculation. The
parameters of the $h_c$ resonance are determined to be $M(h_c)=3525.31\pm0.11$\,MeV/$c^2$ and
$\Gamma(h_c)=0.70\pm0.28$\,MeV, where the errors are statistical only.

The MC-determined selection efficiency $\epsilon_i$ and
yield $N_i$ for each $\eta_c$ decay mode are listed in Table~\ref{tb:yields}.
\begin{table}
\caption{MC-determined efficiencies $\epsilon_i$ and yields $N_i$ for
$\psi(3686)\to\pi^0 h_c, h_c\to\gamma\eta_c, \eta_c\to X_i$, where $X_i$ refers to the 16 final states .}
\label{tb:yields}
\begin{center}
\begin{tabular}{l|c|c }
\hline
\hline
Mode & $\epsilon_i(\%)$ &  $N_i$  \\ \hline

$p \bar{p}$                                    &22.2  & 15.3 $\pm$ 4.5  \\
$\pi^+ \pi^- \pi^+ \pi^-$                      &12.6  & 100.3 $\pm$ 11.3  \\
$K^+ K^- K^+ K^-$                              &6.6   & 6.6 $\pm$ 2.6  \\
$K^+ K^- \pi^+ \pi^-$                          &8.7   & 38.4  $\pm$ 7.0  \\
$p\bar{p}\pi^+ \pi^-$                          &7.8   & 19.0 $\pm$ 5.4  \\
$\pi^+ \pi^- \pi^+ \pi^- \pi^- \pi^-$          &5.4   & 50.5 $\pm$ 9.0  \\
$K^+ K^- \pi^+ \pi^- \pi^- \pi^-$              &2.7   & 10.3 $\pm$ 4.9  \\
$K^+ K^- \pi^0$                                &11.4  & 54.9 $\pm$ 9.2  \\
$p \bar{p} \pi^0$                              &8.9   & 14.4 $\pm$ 4.6  \\
$\ks K^\pm \pi^\mp$                            &8.9   & 107.1 $\pm$ 11.8  \\
$\ks K^\pm \pi^\mp \pi^\pm \pi^\mp$            &3.4   & 43.3 $\pm$ 8.0  \\
$\pi^+ \pi^- \eta$                             &4.3   & 32.9 $\pm$ 6.7  \\
$K^+ K^- \eta$                                 &3.0   & 6.7 $\pm$ 3.2  \\
$\pi^+ \pi^- \pi^+ \pi^- \eta $                &1.9   & 38.6 $\pm$ 7.6  \\
$\pi^+ \pi^- \pi^0 \pi^0 $                     &5.5   & 118.4 $\pm$ 12.8  \\
$\pi^+ \pi^- \pi^+ \pi^- \pi^0 \pi^0$          &2.2   & 175.2 $\pm$ 17.3  \\ \hline
Total                                          &  -     & 831.9$\pm$35.0\\
\hline
\hline
\end{tabular}
\end{center}
\end{table}
Based on these numbers, we can calculate the product branching ratios
$\mathcal{B}_1(\psi(3686)\to\pi^0 h_c)\times \mathcal{B}_2(h_c\to\gamma\eta_c)\times
\mathcal{B}_3(\eta_c\to X_i)$.  The branching ratio for $\eta_c\to X_i$ for each of the
16 final states $X_i$ can then be obtained by combining our measurements with
$\mathcal{B}_1(\psi(3686)\to\pi^0 h_c)\times \mathcal{B}_2(h_c\to\gamma\eta_c) =
(4.36 \pm 0.42) \times 10^{-4}$, the average of two recent measurements by
CLEO~\cite{ref:cleohc08} and BESIII \cite{ref:bes3hc10}.  These branching ratios,
with both statistical and systematic errors, are presented in Section~\ref{sec:sum}.

\subsection{\bf Measurement of {\boldmath $\eta_c$} resonant parameters \label{etac_measurement}}
\bigskip

In addition to determining the $h_c$ resonant parameters, we can
also measure the $\eta_c$ mass and width with the same event sample.
The decay chain $\hc \to \gamma\ec$, $\eta_c \to X_i$ is
reconstructed and kinematically fitted in the 16 $\eta_c$ final
states $X_i$.  For candidates with satisfactory kinematic fits, we
use the resulting track and photon momenta to compute the hadronic
mass.  We populate distributions of this hadronic mass by removing
our previous $E1$ photon-energy and $M(\eta_c)$ requirements and
selecting candidates inside a $\pi^{0}$ recoil mass window of $\pm
5\,\rm MeV/c^2$ around the $h_c$ mass, keeping all other criteria
unchanged.

The line shape for the $\eta_c$ signal for these fits is parameterized as
$(E_{\gamma}^3\times BW(m) \times f_d(E_{\gamma})) \otimes R_i(m)$,
where $BW(m)$ is the Breit-Wigner function for $\eta_c$ as a function
of the invariant mass $m$ of the decay products for each channel,
$E_{\gamma}(m) = \frac{M(h_c)^2 - m^2}{2M(h_c)}$ is the energy of the
transition photon in the rest frame of $h_c$, and $f_d(E_{\gamma})$ is a
function that damps the divergent tail due to the $E_{\gamma}^3$ factor,
which incorporates the energy dependence of the $E1$ matrix element and
the phase-space factor.  $R_i(m)$ is the signal resolution function for the $i$th
decay mode, which is parameterized by double Gaussians to account for the
distorting effects of the kinematic fit and detector smearing.  The damping
function that we use was introduced by the KEDR collaboration
\cite{ref:kedretac}:
$$f_d(E_{\gamma}) = \frac{E_0^2}{E_{\gamma}E_0+(E_{\gamma}-E_0)^2},$$
where $E_0 =  E_{\gamma}(m_{\eta_c})$ is the
$E1$-transition-photon peak energy.
The $\eta_c$-candidate hadronic invariant mass spectra from low and high
side bands in the $h_c$ mass (3500$-$3515,
3535$-$3550\,MeV/$c^2$) are used to obtain the background functions for
the $\eta_c$ mass fit.  To mitigate the effects of bin-to-bin fluctuations, these
side-band mass spectra are smoothed before fitting.  A toy MC study was
performed to test the effect of the smoothing and it was demonstrated to be a
robust procedure that does not systematically distort the fit results.  The
channel-by-channel signal and background normalizations are free parameters
determined by the fit.

We ignore the effect of interference between the signal and
background, which was considered in the previous measurement of
$\psi(3686)\to \gamma\eta_c$~\cite{BESIII:2011ab}, because the
branching ratio of $h_c\to\gamma\eta_c$ is about $50\%$ (branching ratio of
$M1$ transition $\psi(3686)\to \gamma\eta_c$ is about $0.3\%$).
The radiative decay of  $h_c \to \gamma 0^-$ should be the same level of $\psi(3686)\to\gamma0^-$,
in this case, the non-$\eta_c$ intensity in $h_c$ is much smaller than that for
$\psi(3686)\to \gamma\eta_c$.

\begin{figure}[tbhp]
\begin{center}
\epsfig{file=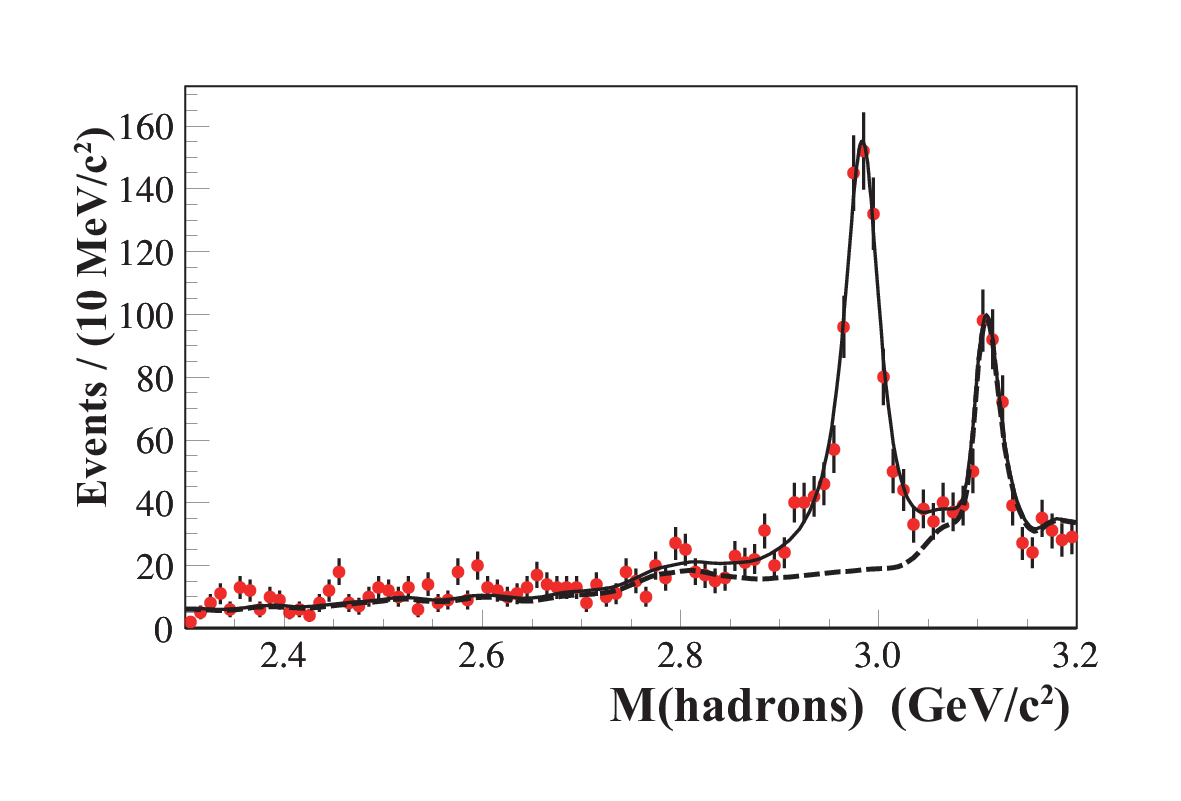, width=14 cm}
\put(-320,180){\bf \large~(a) }\\
\epsfig{file=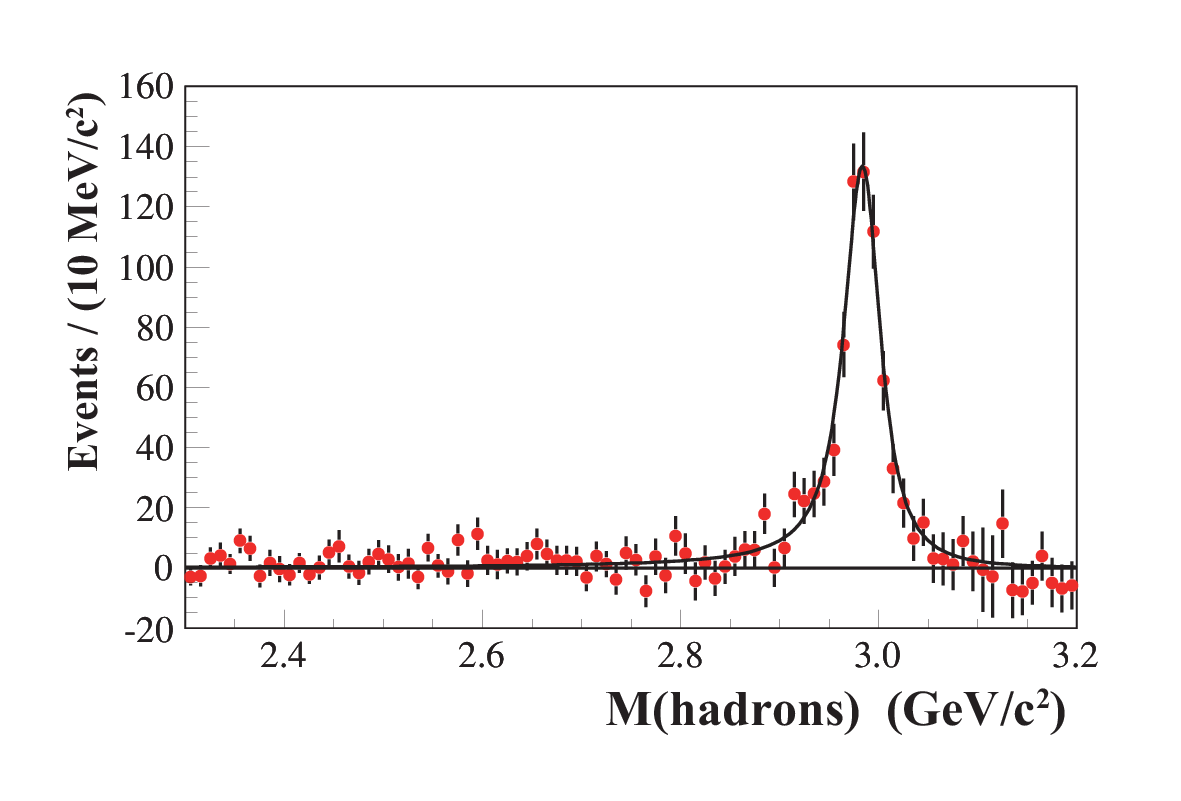, width=14 cm}
\put(-320,180){\bf \large~(b) } \caption{ (a) The hadronic mass
spectrum in $\psi(3686)\to\pi^0{h}_c, {h}_c\to\gamma\eta_c$, $\eta_c
\to X_i$ summed over the 16 final states $X_i$.  The dots with error
bars represent the hadronic mass spectrum in data.  The solid line
shows the total fit function and the dashed line is the background
component of the fit. (b) The background-subtracted hadronic mass
spectrum with the signal shape overlaid. }
 \label{fig:etac_fitting}
\end{center}
\end{figure}

\begin{figure}[tbhp]
\begin{center}
\epsfig{file=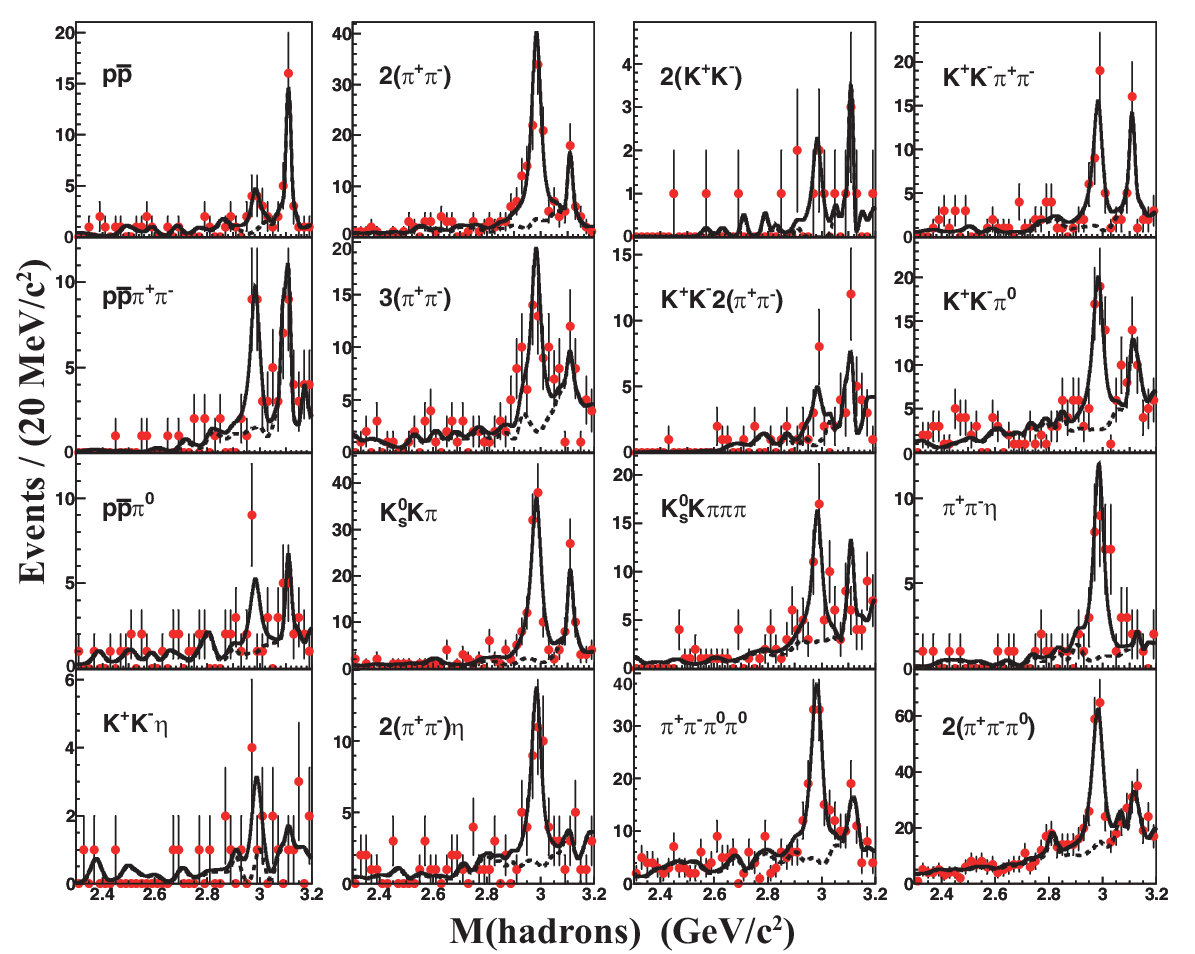,width
= 16 cm} \caption{The simultaneously fitted hadronic mass spectra
for the 16 $\eta_c$ decay channels.}
\label{fig:simfit_hcetac_bkgsub_2.3-3.2}
\end{center}
\end{figure}

Figs.~\ref{fig:etac_fitting} and~\ref{fig:simfit_hcetac_bkgsub_2.3-3.2} show the
hadronic-mass-fit results.  The $\eta_c$ mass and width are
determined to be $M(\eta_c) = 2984.49\pm1.16$\,\rm MeV/$c^2$ and
$\Gamma(\eta_c) = 36.4 \pm 3.2$\,MeV, where the errors are statistical.
The $\chi^2$ per degree of freedom for this fit is 1.52, using the same
$\chi^2$ calculation method to accommodate low-statistics bins as for the
fit to the $\pi^0$ recoil mass spectrum.

\section{\bf Systematic Uncertainties~\label{sec:sys}}
\bigskip

\subsection{{\boldmath $h_c$} parameter measurements}

The systematic uncertainties for the $M(h_c)$ and $\Gamma(h_c)$ measurements are
summarized in Table~\ref{tab:sys1-total}.  All sources are treated as uncorrelated, so
the total systematic uncertainty is obtained by summing them in quadrature. The following
subsections describe the procedures and assumptions that led to these estimates of the uncertainties.

\begin{table}[tbhp]
\centering \caption{ The systematic errors for the $h_c$
mass and width measurements.} \vspace{0.1cm}
\begin{tabular}{l|c|c}
\hline \hline  Sources & $\Delta M_{h_c}$ (MeV$/c^2$) & $\Delta\Gamma_{h_c}$ (MeV) \\
\hline
  Energy calibration             & 0.13   & 0.07   \\
  Signal shape                   & 0.00   & 0.06   \\
  Fitting range                  & 0.04   & 0.16   \\
  Binning                        & 0.02   & 0.01   \\
  Background shape               & 0.01   & 0.08   \\
  Background veto                & 0.01   & 0.08   \\
  Kinematic fit                  & 0.03   & 0.03   \\
  Mass of $\psi(3686)$           & 0.03   & 0.02   \\
 \hline
  Total                          & 0.14   & 0.22   \\
\hline \hline
\end{tabular} \label{tab:sys1-total}
\end{table}

\subsubsection{Energy calibration}\label{sec:caliberr}

The potential inconsistency of the photon-energy measurement between
data and MC is evaluated by studying $\psi(3686)\to\gamma\chi_{c1,2}~(~\chi_{c1,2}\to\gamma
J/\psi,J/\psi\to\mu^+\mu^-$) for photons with low energy and radiative
Bhabha events for photons with high energy.  Discrepancies of $0.4\%$
in the energy scale and $4\%$ in the energy resolution between data and MC
are found.  We vary the photon response accordingly and take the changes
in the results as the estimated systematic error.  For the $M(h_c)$ measurement,
besides the above studies, the reconstructed photon position and error matrix
are taken into account as additional sources of uncertainty.

\subsubsection{Signal shape}

The uncertainty associated with the $h_c$ signal shape in the
$\pi^{0}$ recoil mass spectrum includes contributions from the
photon line shape and the 1-C kinematic fit.  We estimate these by
determining the changes in results after reasonable adjustments in
the photon response. The photon-energy resolution is estimated with
the control sample $\psi(3686)\to\gamma\chi_{c2}$. As above, the
energy resolution in data is found to be about 4\% worse than in the
MC simulation.  We correct for this discrepancy by adding
single-Gaussian smearing to the energy of the $\pi^0$ daughter
photons and then using the alternative $\pi^0$ shape to redo the
fit.  The changes in results are assigned as the systematic errors.

\subsubsection{Fitting range and binning}

The systematic uncertainties due to the fitting of the $\pi^0$ recoil mass spectrum
are evaluated by varying the fitting range and the bin size in the fit.  The spreads of
results obtained with the alternative assumptions are used to assign the systematic errors.

\subsubsection{Background shape}

To estimate the uncertainty associated with the side-band method for assigning
background function shapes, we use an ARGUS function~\cite{Albrecht:1990am}
as an alternative background description for each channel and record the changes in
the fit results.

\subsubsection{Background veto}

The systematic uncertainties associated with the requirements to
suppress background are estimated by varying the excluded ranges.

\subsubsection{Kinematic fit}

Systematic uncertainties caused by the kinematic fit are studied by
tuning the tracking parameters and error matrices of charged tracks
and photons based on the data.  Control samples of $J/\psi\to\phi
f_{0}(980), \phi\to K^+K^-, f_0(980)\to\pi^+\pi^-$, and
$\psi(3686)\to\gamma\chi_{cJ}$ are used for this purpose
\cite{Ablikim:2012pg}.  Channel-by-channel changes of $M(\eta_c)$
and $\Gamma(\eta_c)$ are calculated after the tuning and then
averaged by yields and taken as systematic errors.

\subsubsection{$\psi(3686)$ mass}
The systematic uncertainties of the $M(h_c)$ and $\Gamma(h_c)$ determinations
associated with the uncertainty in the $\psi(3686)$ mass  are estimated to be 0.03\,MeV/$c^2$
and 0.02\,MeV, respectively.  These are found by shifting $M_{\psi(3686)}$ by
one standard deviation according to the PDG value \cite{ref:PDG_2012} and redetermining the
results.

\subsection{{\boldmath $\eta_c$} branching ratio measurements}

The systematic errors in the $\eta_c$ branching ratio measurements
are listed in Tables~\ref{tb:brratioerr1}.
All sources are treated as uncorrelated, so the total systematic
uncertainty is obtained by summing them in quadrature.   The
following subsections describe the procedures and assumptions that
led to the estimates of these uncertainties.

\begin{table}[tbhp]
  \caption{The systematic errors (in \%) in the $\eta_c$ branching ratio measurements of
  the $\eta_c$ exclusive decay channels.} \label{tb:brratioerr1}
  \begin{center}
    \vspace{0.5cm}
    \resizebox{!}{3.0cm}{
    \begin{tabular}{l|cccccccc }
      \hline
      \hline
           Sources &  $p \bar{p}$ &  $2(\pi^+ \pi^-)$ &  $2(K^+ K^-)$ &  $K^+ K^- \pi^+ \pi^-$ &  $p\bar{p}\pi^+ \pi^-$ &  $3(\pi^+ \pi^-)$ &  $K^+ K^- 2(\pi^+ \pi^-)$ &  $K^+ K^- \pi^0$   \\ \hline
           N($\psi(3686)$)     & 4.0 & 4.0 & 4.0 & 4.0 & 4.0 & 4.0  & 4.0   & 4.0      \\
           Tracking            & 4.0 & 8.0 & 8.0 & 8.0 & 8.0 & 12.0 & 12.0  & 4.0      \\
           PID ($K^0_S$)       & 2.0 & 6.0 & 6.0 & 4.0 & 4.0 & 8.0  & 8.0   & 2.0      \\
           Photon eff          & 3.0 & 3.0 & 3.0 & 3.0 & 3.0 & 3.0  & 3.0   & 5.0      \\
           Fit range           & 2.2 & 1.2 & 2.6 & 2.9 & 1.5 & 5.3  & 3.3   & 2.7      \\
           Bkg shape           & 10.3 & 2.5 & 4.7 & 0.9 & 0.3 & 0.2 & 3.5   & 2.8      \\
           Signal shape        & 2.3 & 2.3 & 2.3 & 2.3 & 2.3 & 2.3  & 2.3   & 2.3      \\
           KmFit eff.          & 7.0 & 6.3 & 7.0 & 8.8 & 10.8 & 7.3 & 4.2   & 2.0      \\
           Bkg veto            & 5.9 & 5.5 & 1.1 & 0.6 & 3.1 & 2.3  & 5.2   & 1.7      \\
           Cross feed          & 0.0 & 2.5 & 0.0 & 0.0 & 0.0 & 0.0  & 0.0   & 1.4      \\
           $\eta_c$ decay models& 0.0 & 2.1 & 3.7 & 0.6 & 2.5 & 0.0  & 3.0   & 4.6      \\
           $\eta_c$ line shape & 0.7 & 0.8 & 0.6 & 0.9 & 0.6 & 0.6  & 0.6   & 0.7      \\ \hline
           Sum                 & 15.7 & 14.8 & 14.9 & 14.1 & 15.7 & 18.0 & 17.8 & 10.6       \\ \hline \hline
          \end{tabular} \label{tab:sys2-total}
         }
  \end{center}
\end{table}

\begin{table*}[tbhp]
  \begin{center}
    \vspace{0.5cm}
    \resizebox{!}{3.0cm}{
    \begin{tabular}{l|cccccccc }
      \hline
      \hline
           Sources &  $p \bar{p} \pi^0$ &  $\ks K^\pm \pi^\mp$ &  $\ks K^\pm \pi^\mp \pi^\pm \pi^\mp$ &  $\pi^+ \pi^- \eta$ &  $K^+ K^- \eta$ &  $ 2(\pi^+ \pi^-) \eta$ &  $\pi^+ \pi^- \pi^0 \pi^0 $ &  $2(\pi^+ \pi^- \pi^0)$ \\ \hline
           N($\psi(3686)$)     & 4.0 & 4.0 & 4.0  & 4.0  & 4.0 & 4.0    & 4.0   & 4.0      \\
           Tracking            & 4.0 & 8.0 & 12.0 & 4.0  & 4.0 & 8.0    & 4.0   & 8.0      \\
           PID ($K^0_S$)       & 2.0 & 1.0 & 1.0  & 0.0  & 2.0 & 0.0    & 0.0   & 0.0      \\
           Photon eff          & 5.0 & 3.0 & 3.0  & 5.0  & 5.0 & 5.0    & 7.0   & 7.0      \\
           Fit range           & 7.7 & 2.1 & 1.5  & 0.6  & 6.0 & 1.8    & 0.6   & 2.0      \\
           Bkg shape           & 0.1 & 4.7 & 4.7  & 0.1  & 5.9 & 0.8    & 3.3   & 1.6      \\
           Signal shape        & 2.3 & 2.3 & 2.3  & 2.3  & 2.3 & 2.3    & 2.3   & 2.3      \\
           KmFit eff.          & 6.8 & 6.8 & 7.3  & 2.0  & 1.2 & 6.7    & 2.4   & 2.4      \\
           Bkg veto            & 3.7 & 0.7 & 2.8  & 11.8 & 5.4 & 14.7   & 12.8  & 5.5      \\
           Cross feed          & 0.0 & 0.0 & 0.0  & 0.0  & 0.0 & 0.0    & 1.3   & 0.0      \\
           $\eta_c$ decay models& 5.8 & 2.5 & 5.2  & 5.5  & 8.1 & 0.0    & 0.1   & 0.5      \\
           $\eta_c$ line shape & 0.6 & 0.6 & 0.8  & 0.8  & 0.7 & 0.7    & 0.7   & 0.7      \\ \hline
           Sum                 & 14.8 & 13.2 & 17.0 & 15.4 & 15.3 & 19.4 & 16.4 & 13.3       \\ \hline \hline
          \end{tabular} \label{tab:sys2-total}
         }
  \end{center}
\end{table*}

\subsubsection{Tracking and photon detection}

The uncertainty in the tracking efficiency is 2\% per track and the
uncertainty due to photon detection is 1\% per photon~\cite{Ablikim:2011kv}.
MC studies demonstrate that the trigger efficiency for signal events is almost 100\%,
so that the associated uncertainty in the results is negligible.

\subsubsection{PID and $K^0_S$ reconstruction}

The systematic uncertainties due to kaon and pion identifications
are determined to be 2\% in Ref.~\cite{Ablikim:2011kv}.  We choose
$J/\psi\to K^{*0} K^0_S, K^{*0}\to K\pi$  to evaluate the efficiency
of $K^0_S$ reconstruction. The $1\%$ difference between data and MC
is assigned as the systematic error due to this source.

\subsubsection{Kinematic fitting}

The systematic errors associated with kinematic fitting are estimated by
using the control samples of $\psi(3686)\to\pi^0\pi^0 J/\psi$
with $J/\psi$ decay to hadronic final states, which have similar
event topology as $\psi(3686)\to\pi^0 h_c, h_c\to\gamma\eta_c$. The average efficiency
difference between data and MC, with the same $\chi^2$ requirements in the $h_c$
selection, is taken as the systematic uncertainty.

\subsubsection{Cross-feed}

To evaluate the effect of cross-feed among the 16 signal modes, we use samples of
50,000 MC events per mode. We find that $\eta_c \to 2(\pi^+\pi^-),\eta_c \to
K^+K^- \pi^0$ and $\eta_c \to \pi^+\pi^-\pi^0\pi^0$ are contaminated by
$\eta_c\to K^{0}_{S}K^{\pm}\pi^{\mp}$ with levels of $2.5\%$, $1.4\%$, and $1.3\%$, respectively.
These numbers are assigned as the systematic errors associated with cross-feed.
For other channels, this contamination is found to be negligible.

\subsubsection{$\eta_c$ decay models}

\begin{figure}[tbhp]
\begin{center}
\epsfig{file=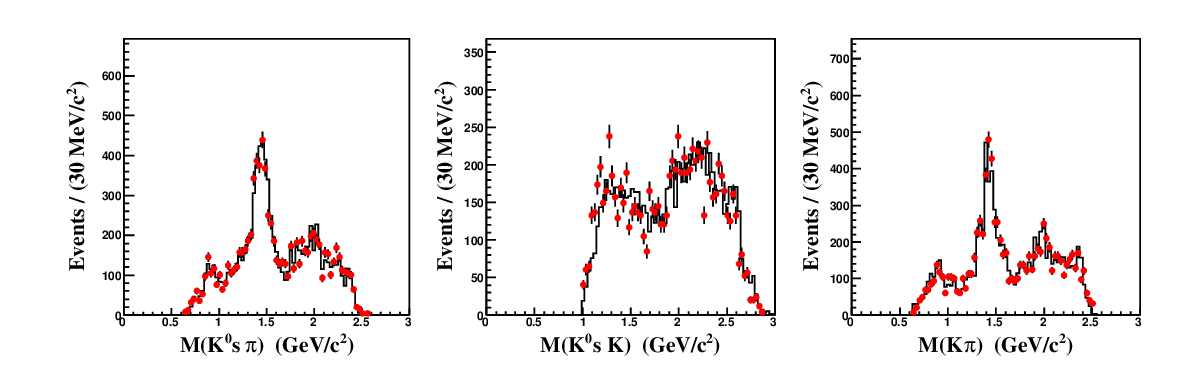,height = 5 cm} \caption{The
dots show the mass spectra for $\psi(3686)\to\gamma\eta_c, \eta_c\to
K^0_{S}K^{\pm}\pi^{\mp}$ in data, and the solid lines are the
corresponding mass spectra from the MC simulation.
\label{fig:reson_kskp}}
\end{center}
\end{figure}

We use phase space to simulate $\eta_c$ decays in our analysis.
To estimate the systematic uncertainty due to neglecting intermediate
states in these decays, we extract invariant masses of $\eta_c$ daughter particles
from $\psi(3686)\to\gamma\eta_c,\eta_c \to X_i$.  We analyze MC
samples generated according to these invariant masses.  To illustrate,
Fig.~\ref{fig:reson_kskp} shows the invariant-mass distribution comparison
between the data and MC for the decay mode $\eta_c\to K^0_{S}K^{\pm}\pi^{\mp}$. In addition,
for channels with low statistics and well-understood intermediate states, MC samples
with these intermediate states were generated according to the relative branching
ratios given by PDG. The spreads of the efficiencies obtained from the phase-space
and alternative MC are taken as the systematic errors.

\subsubsection {$\eta_c$ line shape}

Because of the $\eta_c$ mass window  requirement in our event
selection, the line shape of $\eta_c$ could be a source of
systematic error in the measurement. We vary the input $\eta_c$
resonant parameters by one standard deviation to estimate the
uncertainty due to this source.

\subsection{{\boldmath $\eta_c$} parameter measurements \label{etac_systematic_error}}
\bigskip

Systematic errors for the $M(\eta_c)$ and $\Gamma(\eta_c)$
measurements are summarized in Table~\ref{tab:etacsystote1}.  All
sources are treated as uncorrelated, so the total systematic
uncertainty is obtained by summing in quadrature.  The following
subsections describe the procedures and assumptions that led to the
estimates of these uncertainties.

\begin{table}[tbhp]
  \caption{The systematic errors for $\eta_c$ parameter measurements. } \label{tab:etacsystote1}
  \begin{center}
    {
    \begin{tabular}{l|c|c }
      \hline \hline
          Sources                                    & M($\eta_c$) (MeV$/c^2$)              &  $\Gamma(\eta_c)$ (MeV) \\ \hline
          Background shape                          & $0.36$                      &  $1.45$  \\
          Fitting range                             & $0.03$                      &  $0.33$  \\
          Resolution description                    & $0.10$                      &  $0.02$  \\
          Mass-dependent efficiencies               & $0.11$                      &  $0.27$  \\
          Mass-dependent resolutions                & $0.00$                      &  $0.01$  \\
          Kinematic fitting                         & $0.33$                      &  $0.76$  \\
          Fitting method                            & $0.11$                      &  $0.40$  \\ \hline
          Sum                                       & $0.52$                      &  $1.74$  \\ \hline \hline
   \end{tabular} \label{tab:sys2_b-total}
}
  \end{center}
\end{table}

\subsubsection{Background shape}

Our standard background shape is the smoothed $h_c$ side-band shape.
To estimate the systematic uncertainty due to the background
procedure, we change the smoothing level and technique, and vary the
$h_c$ side-band ranges.   The largest changes in results among these
alternatives are assigned as the systematic errors.

\subsubsection {Fitting range}

The systematic uncertainties due to the fitting range are estimated
by considering several alternatives to the standard fitting range of
2.3-3.2\,$\rm GeV/c^2$, 2.4-3.2\,$\rm GeV/c^2$, 2.5-3.2\,$\rm
GeV/c^2$, 2.6-3.2\,$\rm GeV/c^2$, and 2.3-3.15\,$\rm GeV/c^2$.  The
systematic uncertainties are assigned to be the largest differences
between the standard fit results and those from the alternative
ranges.

\subsubsection{Resolution description}

In order to estimate the systematic uncertainties associated with the detector-resolution description, we use
MC signal shapes obtained by setting the $\eta_c$ width to zero as alternatives to double Gaussians.  The
changes in fit results between these two methods provide the systematic errors.

\subsubsection{Mass-dependent efficiency and resolution}

Since the $\eta_c$ signal spreads over a sizable mass range, the uncertainties due to
the use of mass-independent efficiencies and resolutions need to be estimated.
Mass-dependent efficiencies and resolutions are determined from MC simulation and used as an
alternative to the default assumption, and the resulting differences are taken to be the
systematic errors.

\subsubsection{Kinematic fitting}

The method to evaluate the systematic errors due to the kinematic fitting procedure and
momentum measurement is the same as that in the measurement of the $h_c$ parameters.

\subsubsection{Fitting method}

Because we use the smoothed side-band shape to describe the background,
the potential for bias due to the smoothing technique must be considered.
This was investigated with a toy MC study.  We start with a signal sample for each
of the 16 channels selected from our standard MC to have the same statistics as data.
A corresponding background sample for each channel is constructed from the mass
side bands in data.  The hadronic-mass distributions for these samples are then treated
with a variety of smoothing procedures and fitted.  The ranges in the fit results are used
to set the systematic errors from this source.

\section{\bf SUMMARY AND DISCUSSION} \label{sec:sum}
\bigskip

In summary, we have studied the process $\psi(3686)\to\pi^0 h_c$
followed by $h_c\to\gamma \eta_c$ with an exclusive-reconstruction
technique. Using a sample of 106 million $\psi(3686)$ decays we have
obtained new measurements of the mass and width of the $h_c$ and
$\eta_c$ charmonium resonances, and of the branching ratios for 16
exclusive $\eta_c$ hadronic decay modes.

The total yield of events, measured by fitting the $\pi^0$ recoil mass spectrum, is
$832 \pm35$ events, where the error is statistical only.  With these events we measure the mass and width of the $h_c$:
$$\displaystyle M(h_c)=3525.31\pm0.11\pm0.14\,\rm{MeV}/c^2, ~and$$
$$\displaystyle \Gamma(h_c)=0.70\pm0.28\pm0.22\,\rm{MeV},$$
where the first errors are statistical and the second are systematic.
These results are consistent with the results of a previous inclusive measurement  by BESIII~\cite{ref:bes3hc10}:
$$\displaystyle M(h_c)=3525.40\pm0.13\pm0.18\,\rm{MeV}/c^2, ~and$$
$$\displaystyle \Gamma(h_c)<1.44\,\rm{MeV}~(at~90\%~confidence~level).$$

The branching-ratio results
$\mathcal{B}_{1}(\psi(3686)\to\pi^0h_c)\times
\mathcal{B}_{2}(h_c\to\gamma\eta_c)\times\mathcal{B}_{3}(\eta_c\to
X_i)$ and $\mathcal{B}_{3}(\eta_c\to X_i)$
are given in
Table~\ref{tb:resultbrratio}, quoted with the statistical and systematic errors of this measurement
and, for $\mathcal{B}_3$, an additional systematic error associated with the input
branching-ratio product $\mathcal{B}_{1}(\psi(3686)\to\pi^0h_c)\times
\mathcal{B}_{2}(h_c\to\gamma\eta_c)$. Most of our $\mathcal{B}_{3}(\eta_c\to X_i)$
branching-fraction results are consistent with
PDG values~\cite{ref:PDG_2012}, and several branching fractions are measured for
the first time.

\begin{table}[tbhp]
\caption{$\mathcal{B}_1(\psi(3686)\to\pi^0 h_c)\times
\mathcal{B}_2(h_c\to\gamma\eta_c)\times \mathcal{B}_3(\eta_c\to X_i)$ and
$\mathcal{B}_3(\eta_c\to X_i)$ with systematic errors. The third errors in
$\mathcal{B}_3$ measurement are systematic errors due to uncertainty of $\mathcal{B}_{1}(\psi(3686)\to\pi^0h_c)\times\mathcal{B}_{2}(h_c\to\gamma\eta_c)$.}
\label{tb:resultbrratio}
\begin{center}
    \resizebox{!}{3.95cm}{
\begin{tabular}{l|c|c|c }
\hline \hline $X_i$ & $\mathcal{B}_1\times \mathcal{B}_2\times \mathcal{B}_3~(\times 10^{-6})$  & $\mathcal{B}_3~ (\%)$ &$\mathcal{B}_3$ in PDG $(\%)$\\
\hline
$p \bar{p}$ &                                             0.65  $\pm$ 0.19 $\pm$ 0.10  & 0.15$\pm$0.04$\pm$0.02$\pm$0.01 & 0.141$\pm$0.017\\
$\pi^+ \pi^- \pi^+ \pi^-$ &                               7.51  $\pm$ 0.85 $\pm$ 1.11  & 1.72$\pm$0.19$\pm$0.25$\pm$0.17 & 0.86$\pm$0.13\\
$K^+ K^- K^+ K^-$ &                                       0.94  $\pm$ 0.37 $\pm$ 0.14  & 0.22$\pm$0.08$\pm$0.03$\pm$0.02 & 0.134$\pm$0.032\\
$K^+ K^- \pi^+ \pi^-$ &                                   4.16  $\pm$ 0.76 $\pm$ 0.59  & 0.95$\pm$0.17$\pm$0.13$\pm$0.09 & 0.61$\pm$0.12\\
$p\bar{p}\pi^+ \pi^-$ &                                   2.30  $\pm$ 0.65 $\pm$ 0.36  & 0.53$\pm$0.15$\pm$0.08$\pm$0.05 & \textless 1.2~(at~90\%~C.L.)\\
$\pi^+ \pi^- \pi^+ \pi^- \pi^+ \pi^-$ &                   8.82  $\pm$ 1.57 $\pm$ 1.59  & 2.02$\pm$0.36$\pm$0.36$\pm$0.19 & 1.5$\pm$0.50\\
$K^+ K^- \pi^+ \pi^- \pi^- \pi^-$ &                       3.60  $\pm$ 1.71 $\pm$ 0.64  & 0.83$\pm$0.39$\pm$0.15$\pm$0.08 & 0.71$\pm$0.29\\
$K^+ K^- \pi^0$ &                                         4.54  $\pm$ 0.76 $\pm$ 0.48  & 1.04$\pm$0.17$\pm$0.11$\pm$0.10 & 1.2$\pm$0.1\\
$p \bar{p} \pi^0$ &                                       1.53  $\pm$ 0.49 $\pm$ 0.23  & 0.35$\pm$0.11$\pm$0.05$\pm$0.03 & --\\
$\ks K^\pm \pi^\mp$ &                                     11.35 $\pm$ 1.25 $\pm$ 1.50  & 2.60$\pm$0.29$\pm$0.34$\pm$0.25 & 2.4$\pm$0.2\\
$\ks K^\pm \pi^\mp \pi^\pm \pi^\mp$ &                     12.01 $\pm$ 2.22 $\pm$ 2.04  & 2.75$\pm$0.51$\pm$0.47$\pm$0.27 & --\\
$\pi^+ \pi^- \eta$ &                                      7.22  $\pm$ 1.47 $\pm$ 1.11  & 1.66$\pm$0.34$\pm$0.26$\pm$0.16 & 4.9$\pm$1.8\\
$K^+ K^- \eta$ &                                          2.11  $\pm$ 1.01 $\pm$ 0.32  & 0.48$\pm$0.23$\pm$0.07$\pm$0.05 & \textless 1.5~(at~90\%~C.L.)\\
$\pi^+ \pi^- \pi^+ \pi^- \eta $ &                         19.17 $\pm$ 3.77 $\pm$ 3.72  & 4.40$\pm$0.86$\pm$0.85$\pm$0.42 & --\\
$\pi^+ \pi^- \pi^0 \pi^0 $ &                              20.31 $\pm$ 2.20 $\pm$ 3.33  & 4.66$\pm$0.50$\pm$0.76$\pm$0.45 & --\\
$\pi^+ \pi^- \pi^+ \pi^- \pi^0 \pi^0$ &                   75.13 $\pm$ 7.42 $\pm$ 9.99  & 17.23$\pm$1.70$\pm$2.29$\pm$1.66 & --\\ \hline \hline
\end{tabular}
}
\end{center}
\end{table}

Combining our measurement of $M(h_c)$ with the previously-determined mass of the centroid of
the $^3P_{J}$ states leads to
\begin{equation}\label{deltam}
  \displaystyle \Delta~M_{hf}\equiv\langle M(1^3P)\rangle-
  M(1^1P_1)=-0.01\pm0.11~(\rm stat.)\pm0.15~(\rm syst.)\,\rm{MeV/c^2},
\end{equation}
\noindent consistent with the lowest-order expectation that the $1P$ hyperfine splitting is zero.

The line shape of $\eta_c$ was also studied from the $E1$ transition $h_c\to\gamma\eta_c$, and the measured
resonant parameters are:

$$\displaystyle M(\eta_c) = 2984.49 \pm 1.16 \pm0.52 \,\rm{MeV/c^2}, and$$
$$\displaystyle \Gamma(\eta_c) = 36.4 \pm 3.2 \pm 1.7 \,\rm{MeV}.$$

\noindent These results are consistent with the recent BESIII results from
$\psi(3686) \to \gamma \eta_c$~\cite{BESIII:2011ab}:

$$\displaystyle M(\eta_c) = 2984.3 \pm 0.6 \pm0.6 \,\rm{MeV/c^2}, and$$
$$\displaystyle \Gamma(\eta_c) = 32.0 \pm 1.2 \pm 1.0 \,\rm{MeV};$$
and B-factory results from $\gamma\gamma\to\eta_c$ and $B$ decays
~\cite{Vinokurova:2011dy,delAmoSanchez:2011bt}. Because of the
larger $\psi(3686)$ data sample that will be coming from BESIII and
the advantage of negligible interference effects, we expect that
$h_c\to\gamma\eta_c$ will provide the most reliable determinations
of the $\eta_c$ resonant parameters in the future.

\section{\bf ACKNOWLEDGMENTS}
\bigskip

The BESIII collaboration thanks the staff of BEPCII and the
computing center for their hard efforts. This work is supported in
part by the Ministry of Science and Technology of China under
Contract No. 2009CB825200; National Natural Science Foundation of
China (NSFC) under Contracts Nos. 10745001, 10625524, 10821063,
10825524, 10835001, 10935007, 11125525; Joint Funds of the National
Natural Science Foundation of China under Contracts Nos. 11079008,
11179007, 10979058; the Chinese Academy of Sciences (CAS)
Large-Scale Scientific Facility Program; CAS under Contracts Nos.
KJCX2-YW-N29, KJCX2-YW-N45; 100 Talents Program of CAS; Istituto
Nazionale di Fisica Nucleare, Italy; Ministry of Development of
Turkey under Contract No. DPT2006K-120470; U. S. Department of
Energy under Contract Nos. DE-FG02-04ER41291, DE-FG02-91ER40682,
DE-FG02-94ER40823; U.S. National Science Foundation; University of
Groningen (RuG); the Helmholtzzentrum fuer Schwerionenforschung GmbH
(GSI), Darmstadt; and WCU Program of National Research Foundation of
Korea under Contract No. R32-2008-000-10155-0.

\end{document}